\documentclass[prl,aps,floatfix,twocolumn,showpacs,reprint,superscriptaddress]{revtex4-1}

\usepackage[T1]{fontenc}             
\usepackage{mathptmx}                
\usepackage[scaled=.90]{helvet}
\usepackage{graphicx}                
\usepackage{setspace}                
\usepackage[sort&compress]{natbib}   
\usepackage{float}
\usepackage[loose,nice]{units}       
\usepackage{tabularx}                
\usepackage{booktabs}                
\usepackage{amsmath}

\newcommand{\bra}[1]{\ensuremath{\left<#1\right|}}
\newcommand{\ket}[1]{\ensuremath{\left|#1\right>}}
\newcommand{\braket}[2]{\ensuremath{\left<\left.#1\right|#2\right>}}

\newcommand{\vect}[1]{\ensuremath{\mathbf{#1}}}

\newcolumntype{C}{>{\centering\arraybackslash}X}
\newcolumntype{L}{>{\raggedleft\arraybackslash}X}
\newcolumntype{R}{>{\raggedright\arraybackslash}X}

\begin{document}

\title{Nonsequential Two-Photon Double Ionization of Atoms: Identifying the Mechanism}

\author{Morten F{\o}rre}
\email{morten.forre@ift.uib.no}
\affiliation{Department of Physics and Technology, University of Bergen, N-5007 Bergen,
Norway}

\author{S{\o}lve Selst{\o}}
\affiliation{Centre of Mathematics for Applications, University of Oslo, N-0316 Oslo, Norway}

\author{Raymond Nepstad}
\affiliation{Department of Physics and Technology, University of Bergen, N-5007 Bergen,
Norway}

\begin{abstract}
We develop an approximate model for the process of 
direct (nonsequential) two-photon double ionization of atoms. 
Employing the model, we calculate (generalized) total cross sections as well as 
energy-resolved differential cross sections of helium for photon energies ranging from
39 to 54 eV. A comparison with results of \textit{ab initio} calculations 
reveals that the agreement is at a quantitative level. We thus
demonstrate that this complex ionization process 
is fully described by the simple model, providing insight into the underlying
physical mechanism.
Finally, we use the model to calculate generalized cross sections for the
two-photon double ionization of neon  in the nonsequential regime. 
\end{abstract}


\maketitle

Correlated dynamical processes in nature poses unique
challenges to experiments and theory.
A prime example 
of this is the double ionization of helium
by one-photon impact, which has been studied for more than 40 years.
However, it is only during the last 15 years or so, that advances in theory,
modeling and experiment 
have enabled scientists to gain a deeper insight into
the role of electron correlations in this ionization
process~\cite{Briggs,Huetz,Samson1998,Schneider2002,Foumouo2006}.
The corresponding problem of
two-photon double ionization of helium, in the photon energy interval between
39.4 and 54.4 eV, is an outstanding quantum mechanical problem that has been,
and still is, subject to intense research worldwide, both
theoretically~\cite{Pindzola,Feng2003, Bachau2003,
Bachau_EPD,Hu2005,Foumouo2006, Shakeshaft2007, Ivanov2007, Horner2007,
Nikolopoulos2007, Feist2008,Guan2008,Foumouo2008,Palacios2009,Nepstad_PRA_2010}
and experimentally, employing state-of-the-art high-order
harmonic~\cite{Hasegawa_PRA_2005,Nabekawa,Antoine} and free-electron (FEL) light
sources~\cite{Sorokin2007,Rudenko}.  Despite all the  
interest and efforts
that have been put into this research, major fundamental issues remain
unresolved.  What characterizes this particular three-body breakup process is
that the electron correlation is a prerequisite for the process to occur,
i.e., it depends upon the exchange of energy between the outgoing electrons, and
as such it represents a clear departure from an independent-particle picture.

\begin{figure}[ht] 
	\begin{center} 
        \includegraphics[width=8.cm]{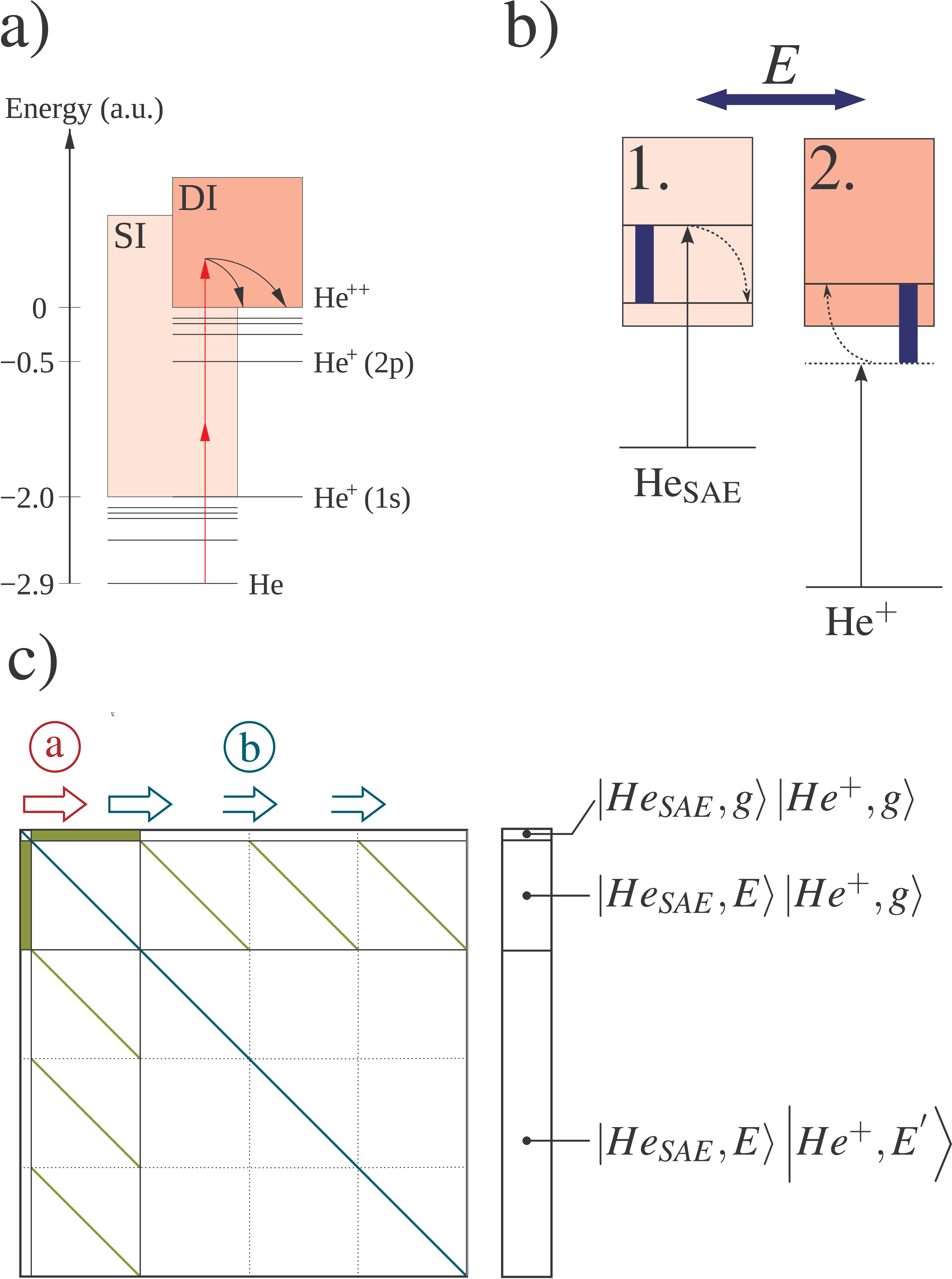}
	\end{center}
	\caption{(color online). a) Sketch of the direct two-photon double ionization 
	process in helium. The abbreviation SI and DI stands for
	single and double ionization continuum, respectively, whereas
	the arrows illustrate the photons that are absorbed by the system.
    b) Sketch of the model process for two-photon double ionization 
    (see text for details). c) Matrix
    representation of the model Hamiltonian, for the case where the outer
    electron is emitted before the inner electron (see text for more details).
	Atomic units (a.u.) are used in the figure (1 a.u. of energy
	corresponds to 27.2 eV).
	}
	\label{fig1} 
\end{figure}

In this Letter, we present a novel approximate model for the direct or
nonsequential two-photon double ionization process in helium, sketched in
Fig.~\ref{fig1}~(a).  We show that the simple model predicts the essential features
of the process, even at a quantitative level, which is quite surprising given
the very high complexity of the problem.  In particular, we find very good agreement
between the model predictions and the results obtained by solving the
time-dependent Schr{\"o}dinger equation from first principles, regarding
(generalized) total cross sections as well as energy-resolved differential
cross sections for the process.  
The proposed model may 
be generalized to account for direct
double ionization processes in multi-electron atoms. 
We demonstrate this by
calculating the generalized cross section for nonsequential two-photon double ionization of
neon. 

Few-photon multiple
ionization of noble gases beyond helium have been studied experimentally in some
detail~\cite{Moshammer_2007,Sorokin2007,Benis_2006,Rudenko}, but to the best of
our knowledge, the cross section for the nonsequential two-photon double
ionization process has not yet been obtained.  Therefore, we hope that our
results will encourage further investigation of nonsequential double ionization
processes in various noble gases.
%
%

Reducing a complex quantum mechanical problem to a simple and transparent model
problem, while retaining the essential physics, is very useful in order to
access the underlying physics~\cite{Lein_2000,Schneider2002,Watson_1997}.  With such a goal in mind,
we will now outline a possible physical mechanism for the nonsequential two-photon double
ionization process in an atom, and then proceed to construct a simple quantum mechanical
model which implements these ideas. 
The idea behind the model is that the electrons are considered to be  
distinguishable particles that  can absorb  one photon each.
However, in order to include the effect of the first emitted electron on the second one,
we impose
the additional but important constraint that the absorption of 
the second photon, by the second electron, can only occur 
after the first photon absorption.
In this way, and according to the principle of conservation of energy, the first electron 
may transfer energy to the second
electron as it is emitted, allowing  for the nonsequential ionization process
to take place.

The starting point of our model is the single-active electron approximation
(SAE) where both electrons are considered to be independent particles and
treated differently in that they are both assumed to move in their respective 
ionization potentials. That is, the 'outer' electron moves in an effective  
potential set up by the nucleus of charge $Ze$ ($e$ is the elementary charge), 
the 'inner' electron and the $Z-2$ other electrons. The inner electron
sees a corresponding screened potential given by the nucleus and 
the $Z-2$ remaining electrons.  
We will label these two different cases simply
by '$A$' and '$B$', respectively.
Following this procedure, the wave function of the ground state
may be approximated by the product ansatz 
\begin{equation}
    \label{eq1}
    \Psi(\vect{r}_{A},\vect{r}_{B})=\psi_{A}(\vect{r}_{A})\psi_{B}(\vect{r}_{B}),
\end{equation}
where $\psi_{A}$ and $\psi_{B}$ refer to the one-electron wave function of 
electron $A$ and $B$, respectively.

Now, the first ionization event in the direct two-photon double ionization process
can be represented by the one-electron dipole coupling between the ground state
wave function of 
\textit{either} $A$ or $B$, 
i.e., the state $\ket{A, E_A^0}$ or $\ket{B, E_B^0}$, 
and their respective continuum states,
$\ket{A, E_A}$ and $\ket{B, E_B}$, where $E_A^0$ and $E_B^0$, and $E_A$ and $E_B$ represent
the energies of the ground and continuum states, respectively.
In the product basis representation~(\ref{eq1}), with the length gauge formulation of
the light-matter interaction, the dipole coupling matrix elements may be
written on the following simple form
(for the case where electron $A$ is assumed to be emitted first),
\begin{equation}
    \label{eq2}
    \bra{A, E_A^0}-e\vect{E}(t)\cdot\vect{r}_A\ket{A, E_A}\braket{B, E_B^0}{B, E_B^0},
\end{equation}
where $\vect{E}(t)$ is the time-dependent electric field that defines the laser
pulse, which is assumed to be linearly polarized along the z-axis.  
These coupling elements are related to the one-photon (one-electron) photoionization cross 
section via the relation~\cite{Cormier_1995}
\begin{equation}
\label{eq3}
\sigma_{A} =  4 \pi^2\alpha \left(E_A - E_A^0\right)  \left| \langle A, E_A^0 | z_A | A, E_A \rangle \right|^2, 
\end{equation}
where $\alpha$ is the fine structure constant.

At the
instant of ionization of electron $A$, electron $B$ remains unaffected.
However, once electron $A$ has absorbed its photon, we allow for the possibility
that electron $B$ (but not $A$) can be hit by a second photon. This secondary
process is included into the model by introducing additional dipole couplings
between the $B$ ground state
and its corresponding one-electron continuum states
in the following way:
\begin{equation}
    \label{eq4}
    \bra{B, E_B^0}-e\vect{E}(t)\cdot\vect{r}_B\ket{B, E_B}\delta(E_A,E^{'}_A)
\end{equation}
Note here that there are only non-vanishing couplings between SAE states (system
$A$) of the same energy, i.e., the resulting coupling matrix attains a very
simple structure, as shown in Fig. \ref{fig1}, with typically only a few hundred
different couplings. 
The same procedure may also be followed with $A$ and $B$
interchanged, however, this will neccessarily yield the same result, 
and therefore need not explicitly be considered.

%
\begin{figure}[t] 
	\begin{center} 
		\includegraphics[width=8.5cm]{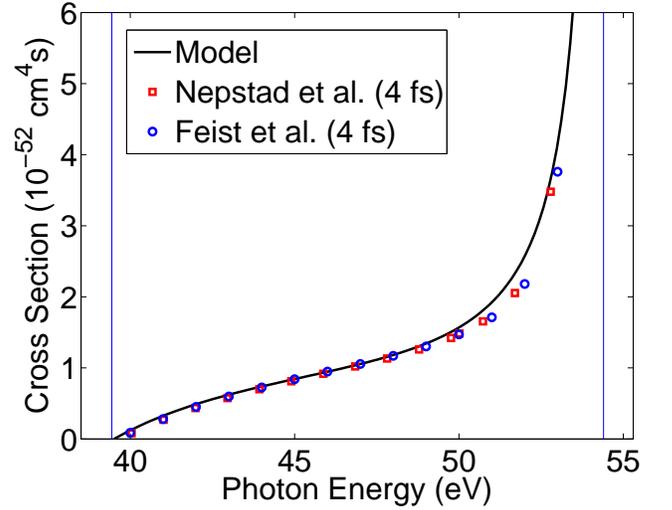} 
	\end{center}
	\caption{(color online). 
Total integrated (generalized) cross section for the nonsequential 
two-photon double ionization of helium.
Black line: present model result obtained with Eq.~(\ref{eq5});
open (blue) circles: {\it ab initio} result of Feist {\it et al.}~\cite{Feist2008}
obtained with a 4 fs pulse;
and open (red) squares: corresponding {\it ab initio} result of Nepstad {\it et al.}~\cite{Nepstad_PRA_2010}.     
The vertical lines define the two-photon direct double ionization region.}
	\label{fig2} 
\end{figure} 

The couplings~(\ref{eq2}) and~(\ref{eq4}) and the mentioned constraints, along
with the corresponding diagonal energies, constitute the entire model that we
propose.  To this end, we would like to add that all excited, bound states
have been left out of the model, as they play no role
in the present context.
As a matter of fact, despite the extremely simple form of the model
matrix elements, with no explicit presence of the correlation potential, it
actually allows for the possibility that the two electrons exchange energy in
the excitation process.  Thus, both electrons may be emitted into the continuum
even though the energy of the secondary photon may not itself be sufficient to
eject  the inner electron into the continuum.

Applying second order perturbation theory to the resulting model Hamiltonian, one
can show that the single-differential cross section for the direct two-photon
double ionization of an atom is simply given by
\begin{eqnarray}
\label{eq5} 
\frac{d \sigma}{d E} & 
= & \frac{1}{2} \left[ f(E)+f\left(2 \hbar \omega + E^0_A+E^0_B-E\right)\right] \\
\nonumber
f(E) & \equiv & \frac{\hbar^3\omega^2}{\pi} \frac{\sigma_{A}\!\left(E-E^0_A\right)\; 
\sigma_{B}\!\left(2 \hbar \omega - E + 
E^0_A\right)}{\left(E-E^0_A\right)\left(2 \hbar \omega - E + E^0_A\right)\left(E-E^0_A-\hbar 
\omega\right)^2},
\end{eqnarray}
where $\sigma_{A}$ and $\sigma_{B}$ now refer 
to the {\it total} one-photon single ionization
cross sections of $A$ and $B$, respectively, $E$ is the excess energy, and where
we have explicitly accounted for the exchange symmetry of
identical particles and the possibility that either the inner or the outer
electron is emitted first.
At this point we would like to emphasize that the only parameters
needed in order to calculate the nonsequential two-photon double ionization cross
section within the model framework, is the
effective binding energies of electron $A$ and $B$, as well as their respective
one-photon single ionization cross sections. For instance, for helium all these 
parameters are well known.
The model may straightforwardly be generalized to account
for e.g. nonsequential three-photon triple ionization processes in atoms.
A more detailed exposition of the model and a derivation of the perturbation theory
expression for the cross section, will be outlined in a forthcoming communication.

%
\begin{figure}[t]
	\begin{center}
                   \includegraphics[width=8.5cm]{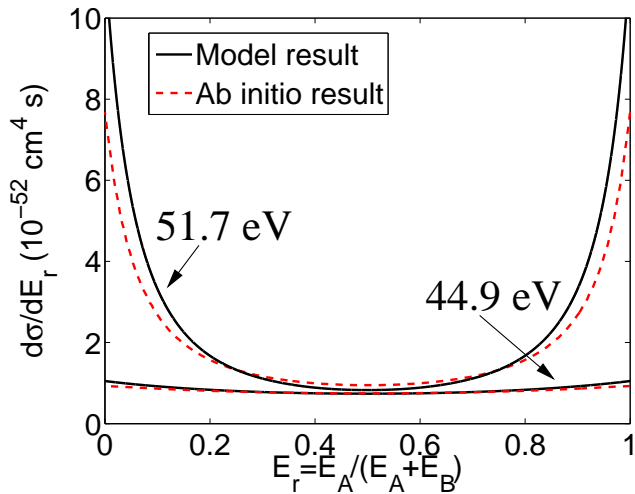}
	\end{center}
	\caption{(color online). 
Electron energy distribution for two-photon double ionization of
helium at photon energies of 44.9 and 51.7 eV.
Solid (black) line: model result; and 
dashed (red) line: {\it ab initio} result.
} 
	\label{fig3} 
\end{figure}
In Fig.~\ref{fig2} we compare the total cross section obtained using the
approximate model, Eq.~(\ref{eq5}), (black line in the figure), 
with the {\it ab initio} result of
Feist {\it et al.}~\cite{Feist2008} (blue circles) and 
Nepstad {\it et al.}~\cite{Nepstad_PRA_2010} (red squares),
both of which were obtained by 
solving the time-dependent
Schr{\"o}dinger equation of helium from first principles. 
The model result is obtained using tabulated values for
the absolute one-photon photoionization cross section of helium,
as obtained experimentally by West and Marr~\cite{West_1976}.
Figure~\ref{fig3} shows corresponding energy-resolved single-differential cross
sections at two selected photon energies, 44.9 and 51.7 eV.  
As a matter of
fact, the agreement between the model result and the {\it ab initio} results 
is almost perfect in Figs.~\ref{fig2} and~\ref{fig3}, in particular for the lower photon energies,
demonstrating the strength of this
extremely simple model in predicting accurate values for the generalized cross
section in direct two-photon double ionization processes. 
Formula~(\ref{eq5}) predicts a sharp rise of the total cross section in the vicinity
of the threshold at 54.4 eV, which is in agreement with 
recent {\it ab initio} calculations~\cite{Horner2007,
Shakeshaft2007,Horner2008,Feist2008,Palacios2009,Nepstad_PRA_2010}.

As mentioned in the introduction, the problem of nonsequential two-photon 
double ionization of helium has been
subject of intense research in recent years, and accurate predictions for the
generalized cross section remain elusive~\cite{Pindzola,Feng2003, Bachau2003,
Bachau_EPD,Hu2005,Foumouo2006, Shakeshaft2007, Ivanov2007, Horner2007,
Nikolopoulos2007, Feist2008,Guan2008,Foumouo2008,Palacios2009,Nepstad_PRA_2010,
Hasegawa_PRA_2005,Nabekawa,Antoine,Sorokin2007,Rudenko} as the values
obtained for the cross section for the reaction may differ by as much as an 
order of magnitude.
On the theoretical side, the great discrepancies that remain between different
approaches are usually ascribed to the different ways electron correlations are
handled in the final state.  To this end, we hope that the predictions of the present model
study may shed new light on this controversy.

\begin{figure}[t]
	\begin{center}
		\includegraphics[width=8.5cm]{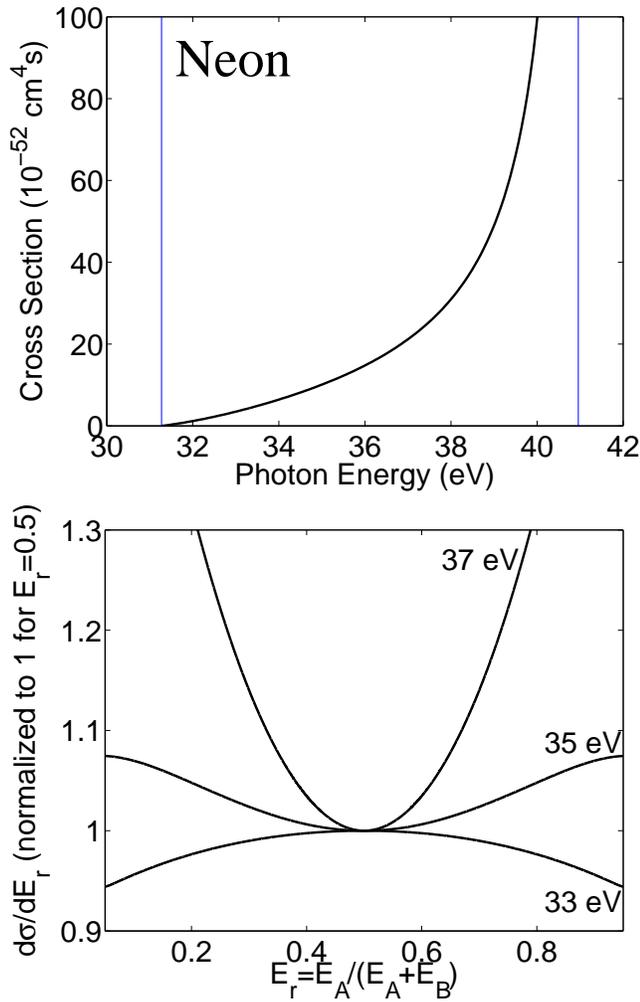}
	\end{center}
	\caption{ 
Upper panel: generalized total cross section for the process of 
two-photon double ionization of neon in the direct regime.
The model result is obtained using Eq.~(\ref{eq5}), inserting 
available experimental values for the
total one-photon single ionization cross sections of
neon~\cite{West_1976} and Ne$^+$~\cite{Covington_2002}, respectively. 	
The vertical lines define the two-photon direct double ionization region.
Lower panel: normalized energy distributions at various photon energies.
	}
	\label{fig4}
\end{figure}

Having justified the validity of our simple approach, we now turn to a
more complex problem, namely the process of nonsequential 
two-photon double ionization of neon.
Inserting, in Eq.~(\ref{eq5}), the correct first and second ionization energies of neon, i.e.,
21.6 and 40.9 eV, as well as experimental values for the photoionization
cross sections of Ne~\cite{West_1976} and Ne$^+$~\cite{Covington_2002}, 
obtained using synchrotron radiation, the resulting
model prediction for the double ionization cross section 
is shown in Fig.~\ref{fig4} (upper panel). The lower panel
shows the corresponding electron energy distribution at three
selected photon energies. Interestingly, at lower photon energies,
the energy distribution exhibits a maximum (negative concavity) when
both electrons are emitted with the same energy, while at higher photon
energies the distribution is U-shaped. 
In sharp contrast to this trend, for helium,
the model yields a U-shaped energy distribution
for all photon energies  (see Fig.~\ref{fig3}).

%

In conclusion, we have implemented an approximate and very simple 
model to study the two-photon double ionization process of helium
in the direct regime, i.e., at photon energies below 54.4 eV where the 
sequential ionization process is energetically inaccessible.
We have investigated the validity of the model by calculating generalized 
total cross sections and energy-resolved differential cross sections 
and compared the model
results with corresponding results obtained by accurate {\it ab initio} calculations. 
Quantitative agreement
between model results and the full results was achieved in all 
considered cases, demonstrating the general validity of the model for the
two-photon double ionization process. 
Finally, we have obtained the cross section for nonsequential 
two-photon double ionization of neon, demonstrating
that the model has a great potential 
to be used in studies of nonsequential
multiphoton multiple ionization processes in more complex atomic systems. 
This is an avenue of research we plan to pursue in the future.

%
%
\begin{acknowledgments}
This work was supported by the Bergen Research Foundation (Norway).
The {\it ab initio} calculations were performed on the 
Cray XT4 (Hexagon) supercomputer installation
at Parallab, University of Bergen (Norway). 
\end{acknowledgments}

\end{document}